\documentclass[symmetry,article,accept,moreauthors,pdftex]{Definitions/mdpi}

\usepackage{siunitx}
\usepackage{aas_macros}
\usepackage{rotating}
\usepackage{braket}
\usepackage{bbm}

\firstpage{1} 
\pubvolume{1}
\issuenum{1}
\articlenumber{0}
\pubyear{2021}
\copyrightyear{2021}
\externaleditor{{Academic Editor: Matthew Mewes} 
} 
\datereceived{1 April 2021} 
\dateaccepted{8 May 2021} 
\datepublished{} 
\hreflink{https://doi.org/} 

\Title{New Constraints on Lorentz Invariance Violation from Combined Linear and Circular Optical Polarimetry of Extragalactic Sources}
\TitleCitation{New Constraints on Lorentz Invariance Violation from Combined Linear and Circular Optical Polarimetry of Extragalactic Sources}

\Author{Roman Gerasimov $^{1,}$*\orcidA{}, Praneet Bhoj $^{1,}$* and Fabian Kislat $^{2,}$*\orcidB{}}
\AuthorNames{Roman Gerasimov, Praneet Bhoj and Fabian Kislat}
\AuthorCitation{Gerasimov, R.; Bhoj, P.; Kislat, F.}

\address{%
$^{1}$ \quad Center for Astrophysics and Space Sciences, University of California, San Diego, La Jolla, CA 92093, USA\\ 
$^{2}$ \quad Department of Physics \& Astronomy and Space Science Center, 
University of New Hampshire, \mbox{Durham, NH 03824, USA}}
\corres{Correspondence: romang@ucsd.edu (R.G.); pnbhoj@ucsd.edu (P.B.); fabian.kislat@unh.edu (F.K.)}

\abstract{Expanding on our prior efforts to search for Lorentz invariance violation (LIV) using the linear optical polarimetry of extragalactic objects, 
 we propose a new method that combines linear and circular polarization measurements. 
While existing work has focused on the tendency of LIV to reduce the linear polarization degree, this new method additionally takes into account the coupling between photon helicities induced by some models.
This coupling can generate circular polarization as light propagates, even if there is no circular polarization at the source.
Combining significant detections of linear polarization of light from extragalactic objects with the absence of the detection of circular polarization in most measurements results in significantly tighter constraints regarding LIV.
The analysis was carried out in the framework of the Standard-Model Extension (SME), an effective field theory framework to describe the low-energy effects of an underlying  fundamental quantum gravity theory.
We evaluate the performance of our method by deriving constraints on the mass dimension $d=4$ CPT-even
 SME coefficients from a small set of archival circular and linear optical polarimetry constraints and compare them to similar constraints derived in previous works with far larger sample sizes and based on linear polarimetry only.
The new method yielded constraints that are an order of magnitude	tighter even for our modest sample size of 21 objects.
Based on the demonstrated gain in constraining power from scarce circular data, we advocate for the need for future extragalactic circular polarization surveys.
}

\keyword{Lorentz invariance; Standard-Model Extension; AGN; polarization} 

\begin{document}

\section{Introduction}
\label{section:introduction}
Einstein's theory of general relativity provides an excellent classical model of gravitation, and~the Standard Model of particle physics is a well-established quantum theoretical model of particles and all forces except gravity.
Together, they provide a well-tested description of nature at experimentally attainable energies.
However, at~the Planck scale ($E_P \approx \SI{1.22e19}{GeV}$), a~quantum-consistent theory of gravity is required.
A lack of direct experimental input to guide the development of the theory poses a significant challenge.
Additionally, the~failure of the Large Hadron Collider to detect evidence regarding physics beyond the Standard Model, including supersymmetry, presents a challenge to many candidate theories~\cite{rpp} (Indications of 
beyond-the-standard-model physics at the $3.1\sigma$ level in B quark decays were presented by the LHCb collaboration after completion of this work~\cite{aaij_2021}).

\textls[-5]{Several theories that attempt to unify gravity and the Standard Model at the Planck scale suggest that there may be deviations from Lorentz invariance at this energy scale~\cite{myers_pospelov_2003, rizzo_2005, amelino_camelia_2015, kostelecky_samuel_1989, burgess_cline_2002, gambini_pullin_1999, pospelov_shang_2012, li_2009}.
This motivates detailed tests of Lorentz symmetry despite the fact that such deviations are expected to be highly suppressed at energies $E\ll E_P$~\cite{kostelecky_potting_1995, kostelecky_summary}.
Tests of this kind are routinely carried out by high-energy physics experiments~\cite{adamson_2010, adamson_2012, mattingly_2005, aaij_2016, carle_2019}; however, reaching progressively higher energies will eventually become unfeasible.
Astrophysical tests in the photon sector have proven to be particularly powerful because tiny deviations from the speed of light as a result of Lorentz invariance violations (LIV) accumulate when photons propagate over very large distances resulting in potentially observable~effects.}

If Lorentz symmetry is broken, the~phase velocity of light in vacuum may depend on the photon energy, polarization, and~direction of propagation. 
The Standard-Model Extension (SME) is an effective field theory framework that extends the Standard Model of particle physics by introducing new, Lorentz, and CPT violating terms in the Lagrangian, while conserving the charge, energy, and~momentum~\cite{kostelecky_summary,colladay_kostelecky_1,colladay_kostelecky_2,kostelecky_renorm,kostelecky_2004}.
Within this framework, group theory considerations allow a classification of potential quantum gravity models in three broad classes with respect to their Lorentz violating effects: birefringent and non-birefringent CPT-even models as~well as CPT-odd models, all of which result in a birefringent photon dispersion~\cite{kostelecky_summary}.
Non-birefringent models result in a dispersion relation that may depend on the photon energy and propagation direction but~will not exhibit any helicity dependence.
The strongest constraints on models of this kind result from astrophysical time-of-flight measurements of gamma-rays emitted by transient events or variable sources~\cite{kislat15,vasileiou_2013, boggs_2004, aharonian_2008, abramowski_2015, albert_2008, biller_1999, ellis_2006, wei_2017, wei_2021}.

Birefringent CPT-even and CPT-odd models can be constrained  more strongly by polarization measurements because the measurement is essentially that of a phase difference between the two polarization modes rather than photon arrival times~\cite{kostelecky_summary, kislat17}.
Most astrophysical radiation processes result in very low circular polarization; however,~linear polarization can be significant.
Birefringence, then, results in a rotation of the linear polarization direction as photons propagate.
If the strength of this effect depends on energy, an~energy dependence of the polarization direction will be observed, even if the polarization at the source does not depend on energy.

When measuring photon polarization over a broad bandwidth, this rotation results in an effective reduction of the observable polarization fraction.
Hence, any observation of linear polarization of light from a distant object can be used to constrain the magnitude of birefringence due to LIV~\cite{kostelecky_summary}.
Some of the strongest constraints of birefringent LIV models come from X-ray polarization measurements~\cite{kaaret_nature,kostelecky_astro_prl,toma_2012,laurent_2011,stecker_2011}.
However, observations of a single source cannot be used to constrain anisotropic models.
In CPT-even models, the LIV terms result in a coupling between the two helicity states, which means neither linear nor circular polarization of light is preserved during propagation.
The helicity of ${\pm}2$ of the coupling necessarily results in~anisotropy.

Within the SME, models can be classified based on their low-energy behavior described above, as~well as an expansion in terms of the mass dimension $d$ of the corresponding operators, and~an expansion in spherical harmonics describing anisotropic effects.
The terms of the expansion are then characterized by a set of coefficients, which can be constrained by experiment~\cite{datatables}.
In the photon sector, terms of odd mass dimension $d$ represent CPT-odd models and even-$d$ terms represent CPT-even models.
Odd-$d$ models are characterized by a set of complex coefficients with $(d-1)^2$ real components.  
Non-birefringent even-$d$ models are characterized by $(d-1)^2$ real components, and~birefrigent even-$d$ models have $2(d-1)^2-8$ coefficients with the same number of real components~\cite{kostelecky_summary}.

We developed a method to combine polarization measurements from many objects in order to individually constrain the coefficients of a given mass dimension.
By applying this method to a large number of optical polarization measurements, we obtained the strongest constraints on individual coefficients to date~\cite{kislat17,kislat18,andy,datatables}.
In essence, we sample the SME coefficient space and, for each set of coefficients, calculate the likelihood to make all given polarization measurements with the published uncertainties of the linear polarization values.
Here, we improve on our approach for CPT-even models characterized by even-$d$ coefficients by incorporating circular polarization measurements of active galactic nuclei in the optical band.

As we will show in Section~\ref{section:polarimetry}, in~these models, the coupling between left-handed and right-handed circular polarization will dominate the reduction of observable linear polarization in most cases.
Hence, circular polarization measurements provide an important additional constraint resulting in significantly tighter constraints than with linear polarization alone.
In this paper, we extend our method to include circular polarization data.
We then apply the method to a set of $21$ linear and circular polarization measurements of quasars in order to derive new constraints on the 10 birefringent SME coefficients of mass dimension $d=4$.
Our new constraints are an order of magnitude stronger than the existing constraints on the individual $d=4$ photon-sector SME~coefficients.

The remainder of the paper is structured as follows.
In Section~\ref{section:sme}, we briefly describe the photon-sector Lagrangian and photon dispersion relation in the minimal SME, and~then derive the underlying equations of our method in Section~\ref{section:polarimetry}.
Section~\ref{section:source} summarizes the assumptions regarding linear and circular polarization at the source that we make in our analysis.
The dataset is described in Section~\ref{section:dataset}.
We explain the Markov-Chain Monte Carlo method used to sample the SME coefficient space and~give the results of our analysis in Section~\ref{section:results}. We~conclude with a summary in Section~\ref{section:conclusion}.

\section{Photon Sector~SME}
\label{section:sme}
For photons in a vacuum, SME operates by adding two extra terms to the standard Lagrangian of the electromagnetic field, such that the total Lagrangian reads~\cite{kostelecky_summary,kostelecky_astro}:
\begin{equation}
    \mathcal{L}=-\frac{1}{4}F_{\alpha\beta}F^{\alpha\beta}+
    \left[\frac{1}{2}\varepsilon^{\gamma \delta \alpha \beta}A_{\delta}(\hat{k}_{AF})_\gamma  F_{\alpha\beta}-
    \frac{1}{4}F_{\gamma\delta}(\hat{k}_F)^{\gamma\delta\alpha\beta}F_{\alpha\beta}\right]
\end{equation}

\noindent where the added terms are placed in brackets, $F_{\alpha\beta}$ is the field tensor, $A_\alpha$ is the $4$-potential of the field ($F_{\alpha\beta}=\partial_\alpha A_\beta-\partial_\beta A_\alpha$), $\varepsilon^{\gamma \delta \alpha \beta}$ is the Levi-Civita symbol, $(\hat{k}_{AF})_\gamma$ and $(\hat{k}_F)^{\gamma\delta\alpha\beta}$ are the Lorentz invariance-violating operators, 
\begin{equation}
    (\hat{k}_{AF})_\gamma=\sum_{d\ge 3,\ d\in \mathrm{odd}}{\left(k_{AF}^{(d)}\right)_\gamma}^{\lambda_1 ... \lambda_{(d-3)}}\delta_{\lambda_1}...\delta_{\lambda_{(d-3)}}
    \label{eq2}
\end{equation}
\begin{equation}
    (\hat{k}_F)^{\gamma\delta\alpha\beta}=\sum_{d\ge 4,\ d\in \mathrm{even}}\left(k_{F}^{(d)}\right)^{\gamma\delta\alpha\beta\lambda_1 ... \lambda_{(d-4)}}\delta_{\lambda_1}...\delta_{\lambda_{(d-4)}}
    \label{eq3}
\end{equation}

\noindent and the sets of coefficients ${\left(k_{AF}^{(d)}\right)_\gamma}^{\lambda_1 ... \lambda_{(d-3)}}$ and $\left(k_{F}^{(d)}\right)^{\gamma\delta\alpha\beta\lambda_1 ... \lambda_{(d-4)}}$ quantify the effect of the SME. The~coefficients are grouped by the mass dimension of the corresponding term in the Lagrangian, $d$. All coefficients must vanish identically if the standard electromagnetic Lagrangian holds perfectly and no LIV effects are present in the~universe.

The added terms allow for all possible violations of Lorentz invariance in rotations and boosts of the electromagnetic field, while maintaining Lorentz invariance for the inertial frame of the observer and, thereby, ensuring that physics is independent of the chosen system of coordinates~\cite{kostelecky_renorm}. Specifically, non-zero components of $\left(k_{F}^{(d)}\right)^{\gamma\delta\alpha\beta\lambda_1 ... \lambda_{(d-4)}}$ give rise to CPT-even terms in the Lagrangian that preserve the CPT symmetry, while non-zero components of ${\left(k_{AF}^{(d)}\right)_\gamma}^{\lambda_1 ... \lambda_{(d-3)}}$ result in CPT-odd terms that violate both CPT and Lorentz~invariance.

In this work, we follow~\cite{kostelecky_renorm} and further restrict our attention to the so-called \textit{minimal SME} that only contains terms of renormalizable mass dimensions, i.e.,~$d\le 4$. This simplification is motivated by the scarcity of the available circular optical polarimetry of extragalactic sources, which necessitates a theory with the smallest number of free parameters in order to derive reliable constraints. We, however, emphasize that the method presented here is universal and can be easily extended to higher mass dimensions ($d>4$) if provided with a sufficient amount of experimental data from future polarimetric surveys and necessary computational~resources.

Under minimal SME, only the $d=3$ term in Equation~(\ref{eq2}) and $d=4$ term in \mbox{Equation~(\ref{eq3})} remain, resulting in the following Lagrangian:
\begin{equation}
    \mathcal{L}=-\frac{1}{4}F_{\alpha\beta}F^{\alpha\beta}+\left[
    \frac{1}{2}\varepsilon^{\gamma \delta \alpha \beta}A_{\delta}\left(k_{AF}^{(3)}\right)_\gamma  F_{\alpha\beta}-
    \frac{1}{4}F_{\gamma\delta}\left(k_F^{(4)}\right)^{\gamma\delta\alpha\beta}F_{\alpha\beta}\right]
    \label{eq4}
\end{equation}

\noindent where $k_{AF}^{(3)}$ has the units of mass and four independent components, and $k_F^{(4)}$ is dimensionless and has $4^4=256$ components, of~which only $19$ are independent due to the required symmetries: $(k_F^{(4)})_{\alpha\beta\gamma\delta}+(k_F^{(4)})_{\alpha\gamma\delta\beta}+(k_F^{(4)})_{\alpha\delta\beta\gamma}=0$, ${(k_F^{(4)})^{\alpha\beta}}_{\alpha\beta}=0$ and $(k_F^{(4)})_{\alpha\beta\gamma\delta}=-(k_F^{(4)})_{\alpha\beta\delta\gamma}=-(k_F^{(4)})_{\beta\alpha\gamma\delta}=(k_F^{(4)})_{\gamma\delta\alpha\beta}$ \cite{kostelecky_renorm}.

The equations of motion associated with the Lagrangian in Equation~(\ref{eq4}) are the modified Maxwell equations with SME-induced LIV. Two leading order plain wave solutions (eigenmodes) exist with orthogonal {electric field vectors} and a phase shift. Following the conventions in~\cite{kostelecky_astro,kostelecky_summary}, we write out the modified dispersion relations for the two solutions as follows:
\begin{equation}
E=\left(1-\varsigma^0\pm\sqrt{(\varsigma^1)^2+(\varsigma^2)^2+(\varsigma^3)^2}\right)\mathbbm{p}
\label{eq5}
\end{equation}

\noindent where $E$ and $\mathbbm{p}$ are the energy and momentum of the photon (setting $c=\hbar=1$) and $\varsigma^\alpha$ are functions of the direction of arrival at the observer and the wavelength of the photon with the functional forms determined by the components of $k_F^{(4)}$ and $k_{AF}^{(3)}$. Specifically, $\varsigma^\alpha$ are defined such that $\varsigma^0$ depends on nine components of $k_F^{(4)}$; $\varsigma^1$ and $\varsigma^2$ depend on the remaining $10$ components of $k_F^{(4)}$; and $\varsigma^3$ depends only on the components of $k_{AF}^{(3)}$. The~observed electromagnetic wave is a superposition of both~eigenmodes.

A non-zero value of $\varsigma^0$ modifies the speed of propagation ($E/p$) for both eigenmodes equally and independently of the photon wavelength for renormalizable mass dimensions. Since $\varsigma^0$ has no effect on polarization, no constraints on its value can be derived in this study. However, $\varsigma^0$ gains wavelength dependency at $d>4$, in~which case its effect may be detected through \textit{vacuum dispersion} (the dependence of the speed of light on the \mbox{wavelength) \citep{kislat15,arrival_delays,kostelecky_astro}.} Non-zero values of $\varsigma^1$, $\varsigma^2$, and $\varsigma^3$ result in a relative difference in the propagation speed between the two eigenmodes and will, therefore, alter the polarization state of the observed wave in-flight (\textit{vacuum birefringence}).

\section{Optical~Polarimetry}\label{section:polarimetry}
\subsection{Monochromatic~Observations}

It is convenient to describe electromagnetic polarization in the Stokes formalism~\cite{Stokes_description} where each polarization state is uniquely identified by a Stokes vector, $\pmb{s}=(q,u,v)$, where $q$, $u$, and $v$ are the intensity normalized Stokes $Q$, $U$, and $V$ parameters. {In this formalism, $v$ sets the total fraction of circular polarization (between $-1$ and $+1$ with negative and positive fractions corresponding to clockwise
and anticlockwise modes respectively, looking outward such that anticlockwise is due East) and $q$ and $u$ encode the linear polarization fraction, $p$, and~angle, $\psi$, according to the equations below:} 
\begin{equation}
    p=\sqrt{q^2+u^2}
    \label{eq_p_def}
\end{equation}
\begin{equation}
    \psi=\frac{1}{2} \mathrm{atan2}\left(u,q\right)
    \label{eq_psi_def}
\end{equation}

\noindent where $\mathrm{atan2}(y,x)$ is the inverse tangent of $y/x$ in the appropriate quadrant ({$\mathrm{atan2}(y,x)=\mathrm{arg}(x+iy)$ for real $x$, $y$}). Following the {International Astronomical Union} (IAU) convention (\cite{IAU_pol_angle}, comm'n 40.8), the~frame of reference in the Stokes space is defined such that $\psi$ is measured from North due East in the {\textit{International Celestial Reference System}} (ICRS) {described} in~\cite{ICRF,ICRS}. {A purely linear polarization state requires $v=0$ and $p\neq0$.}

The eigenmodes in Equation~(\ref{eq5}) are then given by two antiparallel 
Stokes vectors 
$(\varsigma^1,\varsigma^2,\varsigma^3)$ and $(-\varsigma^1,-\varsigma^2,-\varsigma^3)$~\cite{stokes_antiparallel}. 
Both vectors represent the so-called \textit{birefringence axis}, which we denote with $\pmb{\varsigma}$ and, for~convenience, take to be positive:
\begin{equation}
    \pmb{\varsigma}=(\varsigma^1,\varsigma^2,\varsigma^3).
\end{equation}

\begin{figure}[H]
\includegraphics[width=0.9\columnwidth]{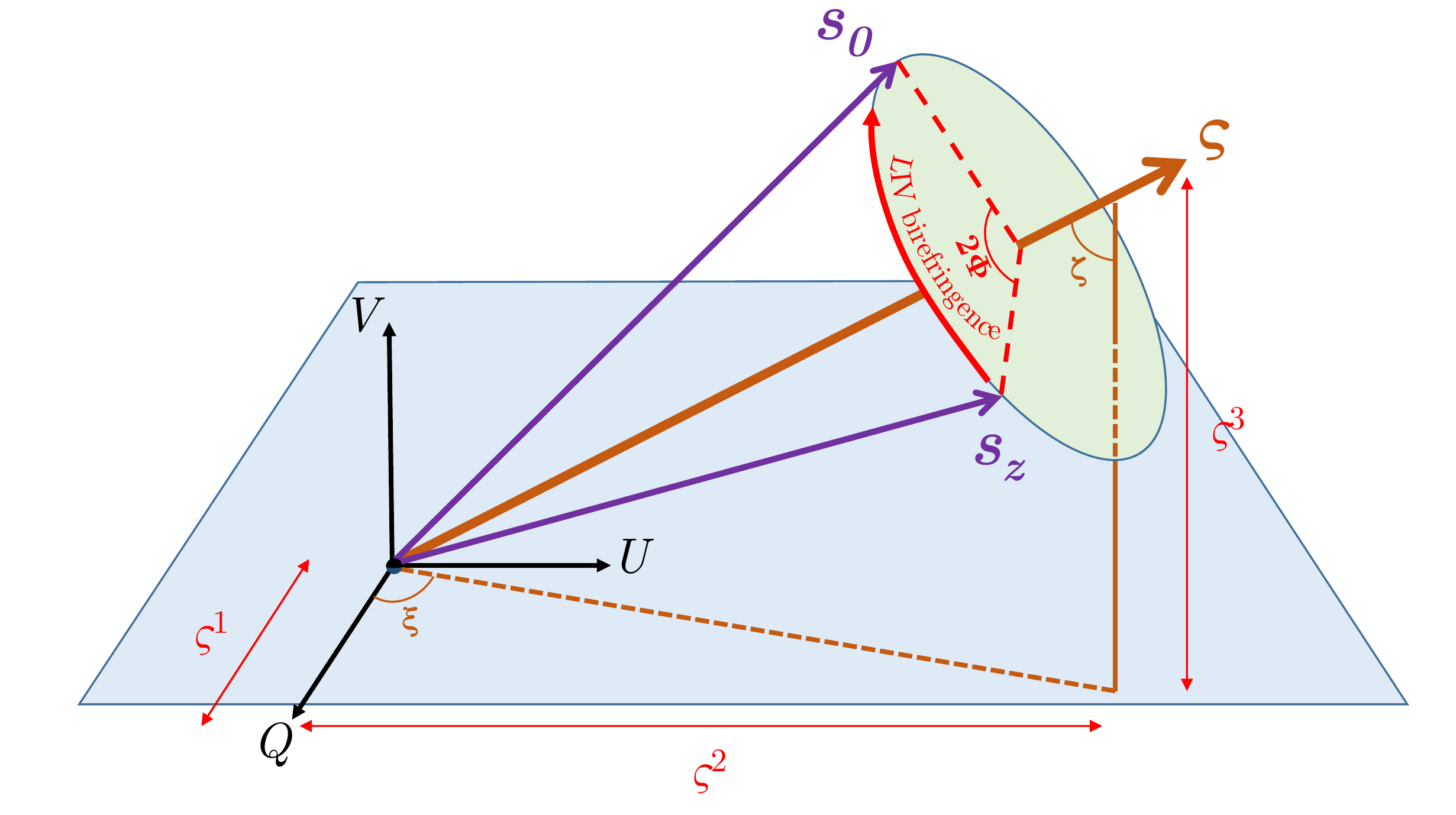}
\caption{Schematic illustration of the Lorentz invariance violation (LIV) effect on the polarization state of a photon in the Standard-Model Extension (SME) framework. The~photon is emitted at the source in an initial polarization state, whose location in the Stokes space (shown here) is indicated with vector $\pmb{s}_z$. In-flight, the~state will precess around the birefringence axis $\pmb{\varsigma}$ through angle $2\Phi$ until, eventually, the~photon arrives at the telescope in the state $\pmb{s}_0$. The~direction of the birefringence axis and the rate of precession are determined by the particular SME configuration. The~blue quadrilateral represents the plane of linear polarization where Stokes $V$ is $0$. Individual components of $\pmb{\varsigma}=(\varsigma^1,\varsigma^2,\varsigma^3)$ and the angles referenced in text ($\xi$, $\zeta$) are labeled as~well.}
\label{fig:diagram}
\end{figure}

The polarization state of a photon (composed of both eigenmodes) propagating in the SME universe, $\pmb{s}$ will then precess around the birefringence axis in uniform circular motion with the revolution period of $\pi/(\omega|\varsigma|)$, {corresponding to the time taken by the phase shift between the two eigenmodes in Equation~(\ref{eq5}) to evolve through $2\pi$}. The~precession direction is against the right hand rule with respect to the birefringence axis. The~process is schematically depicted in Figure~\ref{fig:diagram}. {In the figure, the~intensity-normalized Stokes parameters are presented as Cartesian dimensions in 3D space (so-called Stokes space). The~locus of all purely linear polarization states is represented by a horizontal plane passing through the origin.} Mathematically, the~equation of motion is given by:
\begin{equation}
    \frac{d\pmb{s}}{dt}=2\omega\pmb{\varsigma}\times \pmb{s}=2\omega
    \left(\begin{matrix}
        0 & -\varsigma^{3} & \varsigma^{2} \\
        \varsigma^{3} & 0 & -\varsigma^{1}\\
        -\varsigma^{2} & \varsigma^{1} & 0\\
    \end{matrix}\right) \pmb{s}.
    \label{eq13}
\end{equation}

{In Equation~(\ref{eq13}), the~factor of $2$ arises from the fact that the propagation speed between the two eigenmodes differs by $2|\varsigma|$ as per Equation~(\ref{eq5}). We emphasize that the birefringence axis is defined in Stokes rather than physical space, and, as~such, its existence does not immediately imply a preferred direction in the universe. Instead, the~anisotropy is generated by the directional dependence of the birefringence axis discussed in Section~\ref{section:directional_dependence}.}

The Stokes vector represents the polarization state of a photon arriving from a specific direction in the sky and is, therefore, a function on the celestial sphere, requiring a local reference direction of $\psi=0$ (ICRS North according to the IAU convention).  $v$ is independent of the choice of reference and $q\mp iu$ rotate as spin $\pm2$ objects. 
{In this context, a~function $f$ of celestial coordinates (such as $q$, $u$, $v$, or $q\pm iu$) is said to have spin $s$ if it transforms as $f\rightarrow f\exp(i s \delta\psi)$ when the local North is rotated by $\delta\psi$ radians due East. In~this case, $q\mp iu\rightarrow (q\mp iu)\exp(\pm 2i \delta\psi)$ and $v\rightarrow v\exp(i \delta\psi\times0)$.}

Taking advantage of those symmetries, we re-express $\pmb{\varsigma}$ and $\pmb{s}$ in a {so-called} spin-weighted basis, {such that each dimension of the vector space has a defined spin. In~particular, we follow the convention used in~\cite{kostelecky_summary} and adopt the new basis} as $\pmb{s}=(q-iu,v,q+iu)$ and $\pmb{\varsigma}=(\varsigma^{1}-i \varsigma^{2},\varsigma^{3},\varsigma^{1}+i\varsigma^{2})=(\varsigma^{+},\varsigma^{3},\varsigma^{-})$, where the spins of vector components are $+2$, $0$, and $-2$ respectively. The~transformation into the new basis may be written as:
\begin{equation}
    \hat{S}=
    \left(\begin{matrix}
        1 & -i & 0 \\
        0 & 0 & 1\\
        1 & i & 0 \\
    \end{matrix}\right).
\end{equation}

This, in~turn, re-organizes the equation of motion, Equation~(\ref{eq13}), into:

\begin{equation}
    \frac{d\pmb{s}}{dt}=-2i\omega
    \left(\begin{matrix}
        \varsigma^{3} & -\varsigma^{+} & 0 \\
        -\varsigma^{-}/2 & 0 & \varsigma^{+}/2\\
        0 & \varsigma^{-} & -\varsigma^{3}\\
    \end{matrix}\right)\pmb{s}.
    \label{eq15}
\end{equation}


From this point onward, the~spin-weighted basis in Equation~(\ref{eq15}) will be assumed in all vectors and matrices. Denoting the total precession angle from emission to observation of the photon polarization state in Stokes space around the birefringence axis with $2\Phi$, the~solution to Equation~(\ref{eq15}) is given by the precession matrix obtained using Rodrigues' {rotation} formula  {(\cite{rodriguez_formula}, p. 209)} and transformed into the spin-weighted space with $\hat{S}$:
\begingroup\makeatletter\def\f@size{9}\check@mathfonts
\def\maketag@@@#1{\hbox{\m@th\normalsize\normalfont#1}}%
\begin{equation}
    \begin{split}
    \pmb{s}_0=\hat{S}\left[\hat{S}^{-1}\pmb{s}_z \cos(2\Phi)-(\hat{S}^{-1}\hat{\pmb{\varsigma}}\times\hat{S}^{-1}\pmb{s}_z)\sin(2\Phi)+\hat{S}^{-1}\hat{\pmb{\varsigma}}(\hat{S}^{-1}\hat{\pmb{\varsigma}}\cdot\hat{S}^{-1}\pmb{s}_z)(1-\cos(2\Phi))\right] \\\\
    =\left(\begin{matrix}
        \Upsilon^2 & -2 i e^{-i \xi} \sin(\zeta)\sin(\Phi)\Upsilon & e^{-2 i\xi} \sin^2(\zeta)\sin^2(\Phi) \\
        -i e^{i\xi} \sin(\zeta)\sin(\Phi)\Upsilon & \cos^2(\zeta)+\cos(2\Phi)\sin^2(\zeta) & i e^{-i\xi} \sin(\zeta)\sin(\Phi)\Upsilon^*\\
        e^{2i\xi}\sin^2(\zeta)\sin^2(\Phi) & 2ie^{i\xi} \sin(\zeta)\sin(\Phi)\Upsilon^* & {\Upsilon^*}^2\\
    \end{matrix}\right) \pmb{s}_z
    \end{split}
    \label{eq_precession}
\end{equation}
\endgroup

\noindent where $\hat{\pmb{\varsigma}}$ is the normalized $\pmb{\varsigma}$ and the following definitions are introduced:
\begin{equation}
    \xi=\mathrm{atan2}(i (\varsigma^+ - \varsigma^-),\varsigma^+ + \varsigma^-)=\arg(\varsigma^+)
\end{equation}
\begin{equation}
    \zeta=\mathrm{atan2}(\sqrt{\varsigma^+\varsigma^-},\varsigma^3)
    \label{eq_zeta}
\end{equation}
\begin{equation}
    \Upsilon=\cos(\Phi)+i \cos(\zeta)\sin(\Phi).
\end{equation}

The angles $\xi$ and $\zeta$ have geometric significance and are labeled in Figure~\ref{fig:diagram}. In~Equation~(\ref{eq_precession}), $\pmb{s}_0=(q_0-i u_0,v_0,q_0+i u_0)$ is the observed Stokes vector of the photon on Earth, and $\pmb{s}_z=(q_z-i u_z,v_z,q_z+i u_z)$ is the initial Stokes vector at redshift $z$.

Considering the precession period of $\pi/(\omega|\varsigma|)$ for the polarization vector, the~equation of motion for $\Phi$ reads:
\begin{equation}
    \frac{d\Phi}{dt}=\omega |\varsigma|\ \ \ \implies\ \ \ \Phi=\int_0^{T}\omega |\varsigma| dt=\int_0^{z}\omega_0 |\varsigma| \frac{dz'}{H_{z'}}
    \label{eq_phi}
\end{equation}

\noindent where $T$ is the total time of flight for the photon, $z$ is the redshift of the source, and $\omega_0$ is the observed angular frequency of the photon at $z=0$. We used $dt=-dz/[(1+z)H_z]$ and $\omega(z)=\omega_0 (1+z)$. The~Hubble expansion parameter, $H_z$, is given by
\begin{equation}
    H_z=H_0 \left[ \Omega_r \left(1+z\right)^{4} + \Omega_m \left(1+z\right)^{3} + \Omega_k \left(1+z\right)^{2} + \Omega_\Lambda \right]^{1/2}.
    \label{eq17}
\end{equation}

In Equation~(\ref{eq17}), $H_0$ is the Hubble constant and $\Omega_r$, $\Omega_m$, $\Omega_k$, and $\Omega_\Lambda$ are the present day fractional contributions of radiation, matter, curvature, and dark energy to the energy budget of the universe. In~this study, we adopt $H_0=67.66\ \mathrm{km}\ \mathrm{s}^{-1}\ \mathrm{Mpc}^{-1}=1.4433\times10^{-33}\ \mathrm{eV}$, $\Omega_m=0.3111$, $\Omega_r=9.182\times10^{-5}$, $\Omega_\Lambda=0.6889$, and $\Omega_k=0$, following~\cite{andy,planck_params} (see~\cite{andy} in particular for a discussion of the accuracy of those values in this context and the Hubble tension).

While, in~principle, both CPT-even ($d=4$) and CPT-odd ($d=3$) contributions to the Lagrangian may exist simultaneously, it is generally preferable to begin the search for Lorentz invariance violation with the smallest number of free parameters. We will follow \citep{andy,kislat17,kislat18,kislat15,kostelecky_summary,single_d} and assume that the effects at a specific $d$ dominate. {In particular, we choose to focus on the $d=4$ case as an even mass dimension is required for helicity coupling (for odd $d$, the~birefringence axis is parallel to the Stokes $V$ axis and, therefore, circular polarization of the photon is conserved). Furthermore, since $\varsigma^{3}\propto \omega^{-1}\propto\omega_0^{-1}$ (see Equation~(\ref{eq2})), $\Phi$ is uniquely wavelength-independent in the $d=3$ case, which motivates alternative approaches in~\cite{CMB_1,CMB_2,CMB_3,CMB_4,CMB_5} to $d=3$ searches involving polarimetry of the Cosmic Microwave Background.}

For $d=4$, all components of $k_{AF}^{(3)}$ vanish and, therefore, so does $\varsigma^{3}$. Geometrically, the~$d=4$ case has the birefringence axis, $\pmb{\varsigma}$, constrained to the plane of linear polarization. For, $\varsigma^{3}=0$, $\zeta$ (Equation \ref{eq_zeta}) evaluates to $\pi/2$, allowing the precession formula in \mbox{Equation~(\ref{eq_precession})} to be simplified as follows:
\begin{equation}
    \pmb{s}_0=
    \left(\begin{matrix}
        \cos^2(\Phi) & -i \sin(2 \Phi) e^{-i \xi} & \sin^2(\Phi) e^{-2i\xi} \\
        -\frac{i}{2} \sin(2\Phi)e^{i\xi} & \cos(2\Phi) & \frac{i}{2} \sin(2\Phi)e^{-i\xi} \\
        \sin^2(\Phi)e^{2i\xi} & i \sin(2\Phi)e^{i\xi} & \cos^2(\Phi) \\
    \end{matrix}\right)
    \pmb{s}_z.
    \label{eq_muller_no_V}
\end{equation}

\subsection{Directional~Dependence}\label{section:directional_dependence}

In Equation~(\ref{eq_muller_no_V}), the~dependence of $\Phi$ on the direction of arrival is determined by the particular SME configuration, which we seek to constrain by comparison with observations. Following~\cite{kostelecky_summary,kostelecky_astro,andy}, we limit the dependence to $10$ linearly independent degrees of freedom (corresponding to the $10$ independent components of $k_F^{(4)}$ setting $\varsigma^\pm$) by expanding $\varsigma^\pm$ in terms of $l=2$, $s=\pm 2$ spin-weighted spherical harmonics, $_{s}Y_{l,m}$, where $s$, $l$, and $m$ are the spin, azimuthal, and magnetic numbers, respectively, ($-2\leq m\leq2$):
\begin{equation}
\varsigma^\pm=\sum_{m=-2}^{2}{_{\pm2}Y_{2,m}(\hat{\pmb{n}})(k_{(E)2,m}\pm i k_{(B)2,m}).}
\label{eq6}
\end{equation}

Here, $\hat{\pmb{n}}$ is a unit vector pointing in the direction of photon arrival (opposite to the photon momentum) and $k_{(E,B)l,m}$ are complex coefficients of the expansion that obey the following symmetries:
\begin{equation}
k_{(E,B)l,-m}=(-1)^m \left( k_{(E,B)l,m}\right)^*
\end{equation}

\noindent such that there are $10$ real independent parameters: $k_{(E)2,0}$, $k_{(B)2,0}$, $\mathrm{Re}\left[k_{(E)2,1}\right]$, $\mathrm{Im}\left[k_{(E)2,1}\right]$, $\mathrm{Re}\left[k_{(E)2,2}\right]$, $\mathrm{Im}\left[k_{(E)2,2}\right]$, $\mathrm{Re}\left[k_{(B)2,1}\right]$, $\mathrm{Im}\left[k_{(B)2,1}\right]$, $\mathrm{Re}\left[k_{(B)2,2}\right]$, and  $\mathrm{Im}\left[k_{(B)2,2}\right]$.

{Since none of the terms in Equation~(\ref{eq6}) have spherical symmetry, the~$d=4$ case considered in this study is naturally anisotropic. However, isotropic terms do exist in odd $d$ SME configurations~\cite{kostelecky_summary,kislat15,kislat17,andy}.}

For astronomical tests, $\hat{\pmb{n}}$ may be written in some system of celestial coordinates. In~this work, we use $\hat{\pmb{n}}=(\alpha, \delta)$, where $\alpha$ and $\delta$ are the ICRS~\cite{ICRF,ICRS} right ascension and declination. 

Substituting Equation~(\ref{eq6}) in Equation~(\ref{eq_phi}):
\begin{equation}
    \Phi=\omega_0\left|  \sum_{m=-2}^{2}{_{2}Y_{2,m}(\hat{\pmb{n}})(k_{(E)2,m}+ i k_{(B)2,m})}\right| \int_0^z{\frac{dz'}{H_{z'}}}.
\end{equation}

{$\Phi$ is directly proportional to the comoving distance to the photon source ($\int_0^z dz'/H_{z'}$), which is a unique feature of $d=4$ SME.}

\subsection{Broadband~Observations}

Astronomical observations of the Stokes parameters $q_0$, $u_0$, and $v_0$ are taken over some finite band defined by a dimensionless detection efficiency $\tau(\omega_0)$, rather than a single monochromatic frequency. The~observed effective Stokes parameter, $Q_0$, is then determined by the weighted average of the observed Stokes parameter $q_0$:
\begin{equation}
    Q_0=\frac{\int_0^\infty \tau(\omega_0) q_0(\omega_0) d\omega_0}{\int_0^\infty \tau(\omega_0) d\omega_0}=\mathcal{T}[q_0]
    \label{eq20}
\end{equation}

\noindent where we define the operator $\mathcal{T}$ as the efficiency-weighted average. The~other two Stokes parameters are then $U_0 = \mathcal{T}[u_0]$ and $V_0 = \mathcal{T}[v_0]$. The observed broadband linear polarization fraction ($\Pi_0$) and angle ($\Psi_0$) may then be defined equivalently to Equations~(\ref{eq_p_def}) and~\mbox{(\ref{eq_psi_def})}: 
\begin{equation}
    \Pi_0=\sqrt{Q_0^2+U_0^2}
\end{equation}
\begin{equation}
    \Psi_0=\frac{1}{2} \mathrm{atan2}\left(U_0,Q_0\right).
\end{equation}

For convenience, we introduce a primed system of Stokes coordinates, rotated with respect to the ICRS coordinates anticlockwise through $\xi$:
\begin{equation}
    \pmb{s}'=\left(\begin{matrix}
        q'-i u' \\
        v' \\
        q'+i u' \\
    \end{matrix}\right)=\left(\begin{matrix}
        e^{i\xi} & 0 & 0 \\
        0 & 1 & 0 \\
        0 & 0 & e^{-i\xi} \\
    \end{matrix}\right)\left(\begin{matrix}
        q-i u \\
        v \\
        q+i u \\
    \end{matrix}\right).
\end{equation}

As a result of this rotation, the~polarization angle decreases by $\xi/2$, while both the linear and circular polarization fractions remain unchanged. In~primed coordinates, the~observed broadband Stokes parameters can be expressed in simple forms by assuming that the source polarization spectrum only slowly varies with $\omega$ across the optical range, implying that $\pmb{s}_z(\omega_z)\approx \pmb{s}_z$. This assumption is justifiable by the low redshift (i.e., SME-free) spectropolarimetry~\cite{kislat17} that shows a weak dependence of polarization on wavelength for most galaxies due to the optical range being comparatively narrow. $Q_0$, $U_0$, and~$V_0$ are then related to the Stokes parameters at the source as:
\begin{align}
\label{eq22}
Q_0' &= q_z' \\
\label{eq23}
U_0' &= +u_z' \mathcal{T}[\cos(2\Phi)]+v_z' \mathcal{T}[\sin(2\Phi)] \\
\label{eq24}
V_0' &= - u_z' \mathcal{T}[\sin(2\Phi)]+v_z' \mathcal{T}[\cos(2\Phi)].
\end{align}

The functional forms of $\mathcal{T}[\cos(2\Phi)]$ and $\mathcal{T}[\sin(2\Phi)]$ are entirely determined by the instrument and may be pre-tabulated for a range of $\Phi$ values for fast lookup. Equations~\mbox{(\ref{eq22})--(\ref{eq24})} demonstrate that the total polarization fraction ($\sqrt{q^2+u^2+v^2}$) is conserved in the monochromatic limit ($\mathcal{T}[x]\rightarrow x$); however, it will, in general, decrease with 
redshift for broadband observations due to the polarization washout induced by the frequency dependence of $d=4$ SME effects (i.e., $|\mathcal{T}[x]|<|x|$). 

\begin{figure}[H]
\includegraphics[width=\columnwidth]{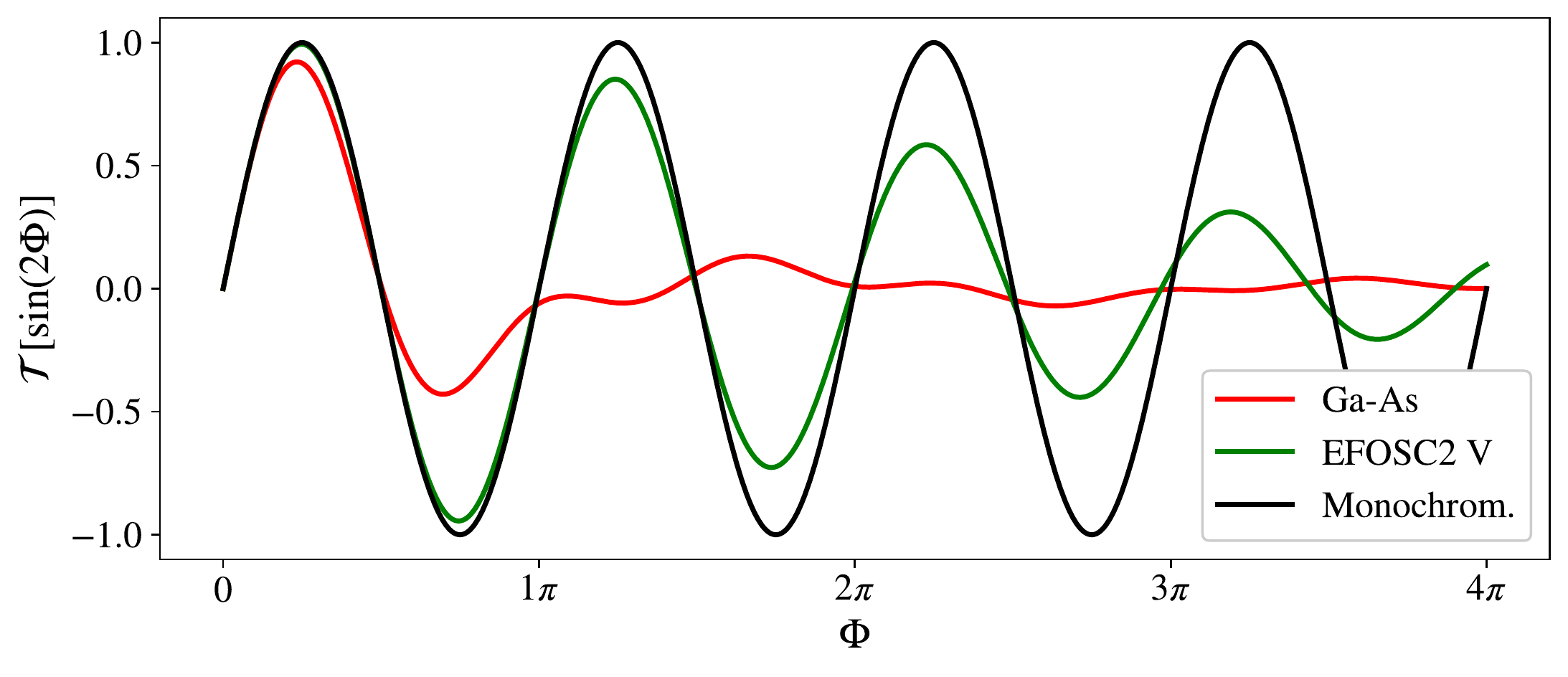}
\caption{The curves show the effect of the broadband operator $\mathcal{T}[x]$ on the amount of LIV-induced linear polarization in-flight (see Equations~(\ref{eq23}) and (\ref{eq24})) as a function of $\Phi$ (see Figure~\ref{fig:diagram}). The~three cases shown correspond to a monochromatic observation (black, $\mathcal{T}[\sin(2\Phi)]\rightarrow\sin(2\Phi)$), a~broadband observation with an unfiltered Ga-As photomultiplier tube (red), and a broadband observation in the \textit{V}-band of the EFOSC2~\cite{EFOSC} instrument (green). This calculation assumes observations at the zenith. See Section~\ref{section:dataset} for more details on instruments, bands, and atmospheric~effects.}
\label{fig:G_integral}
\end{figure}

This effect is illustrated in \mbox{Figure~\ref{fig:G_integral}}, displaying the value of $\mathcal{T}[\sin(2\Phi)]$ as a function of $\Phi$ for the cases of monochromatic, filtered broadband and unfiltered broadband observations. While the amplitude of the operand is conserved in the monochromatic case, it decays in both broadband cases with the unfiltered decay being more rapid due to a wider range of wavelengths~observed.

\unskip

\subsection{Likelihood~Model}

To differentiate SME predictions for the observed broadband polarization parameters ($Q_0$, $U_0$, $V_0$, $\Psi_0$, and $\Pi_0$) derived above from the actual experimental measurements, we introduce subscript $m$ for the latter, such that, in the case of the chosen SME configuration being a perfect description of LIV, we must have $Q_0=Q_m$, $U_0=U_m$ etc. in the absence of other~effects.

In the limit of weak SME motivated by the strict constraints established in previous works~\cite{andy,kislat18,kislat17}, we expect the change in Stokes parameters due to SME to be small, such that the overall ratio of linear to circular polarization ($\Pi/V$) for extragalactic photons is only mildly perturbed in-flight. Since both 
experimental and theoretical considerations (e.g.,~\cite{circ_zero_1,circ_zero_2,circ_zero_3,circ_zero_4}) suggest that linearly polarized emissions dominate over circularly polarized emissions for nearly all realistic extragalactic sources, we further assume that $|v_z|\ll p_z$ and $|V_0|\ll\Pi_0$. 

Under~this assumption, Equations~(\ref{eq23}) and (\ref{eq24}) suggest that the overall effect of the SME is to suppress linear polarization and enhance circular polarization in-flight, as~demonstrated in Figure~\ref{fig:maximization}b where linear and circular polarization fractions are plotted on the same set of axes as functions of redshift. Therefore, SME {coefficients are constrained} by measurements of lower $|V_m|$ and higher $\Pi_m$. In~the extreme case, an~error-free measurement of $V_m=0$ or $\Pi_m=1$ would entirely rule out all SME configurations considered in this~study.

\textls[-35]{Assuming that the source polarization is somehow known (see Section~\ref{section:source}), \mbox{Equations~(\ref{eq22})--(\ref{eq24})} can be used to predict the expected observed polarization for any $d=4$ SME configuration and for any instrument (encoded in the functional form of $\mathcal{T}$). The~prediction must then be compared to experimental results, which may be accomplished by evaluating the likelihood of compatibility of the observed data with theory. For~broadband polarimetry, three types of measurements are~available: }

\begin{itemize}
    \item linear polarization fraction, $\Pi_m$,
    \item polarization angle, $\Psi_m$, and
    \item circular polarization fraction, $V_m$.
\end{itemize}

In our method, we advocate deriving constraints on SME parameters from the linear and circular polarization fraction measurements, $\Pi_m$ and $V_m$. The~polarization angle, $\Psi_m$, is still essential in the method (Section \ref{section:source}), but~will not be used in the likelihood model~directly.

Since the measured circular polarization, $V_m$, may take both positive and negative values, we assume a Gaussian parent distribution with the mean of $V_m$ and the standard deviation of $\sigma_V$. The~likelihood of compatibility of such a measurement with the SME prediction $V_0$ may be written as follows:
\begin{equation}
    \mathcal{P}_\mathrm{circ}=\begin{cases}
        \int_{V_0}^\infty\frac{1}{\sigma_V \sqrt{2 \pi}}\exp\left(-\frac{(V-V_m)^2}{2\sigma_V^2}\right)dV,& \text{if } V_0 \geq 0\\
        \int_{-\infty}^{V_0}\frac{1}{\sigma_V \sqrt{2 \pi}}\exp\left(-\frac{(V-V_m)^2}{2\sigma_V^2}\right)dV,              & \text{if } V_0 \leq 0\\
        1,              & \text{if } V_0 = 0.\\
    \end{cases}
\end{equation}

{The integration bounds over the Gaussian distribution in the equation above are chosen such that a measurement ($V_m$) is considered compatible when its absolute value ($|V_m|$) is as large or larger than the SME prediction ($|V_0|$). In~other words, $\mathcal{P}_\mathrm{circ}$ is the probability of $V_m>V_0$ if $V_0$ is positive and $V_m<V_0$ if $V_0$ is negative.}

The measured linear polarization degree, $\Pi_m$, may display strongly non-Gaussian behavior due to being an intrinsically positive quantity. {Following the derivation in Appendix \ref{section:appendix}}, we assume $\Pi_m$ to be drawn from the Rice distribution ($P(\Pi|\hat{\Pi})$, \mbox{Equation~(\ref{eqA5}))} with $\hat{\Pi}=\Pi_m$ and $\mathrm{Var}[\Pi]=\sigma_\Pi^2$, where $\sigma_\Pi$ is the uncertainty of the measurement and $\mathrm{Var}[\Pi]$ is given in \mbox{Equation~(\ref{eqA7})}. The~statistical distribution of the linear polarization fraction and the bias in linear polarization measurements are discussed in Appendix \ref{section:appendix}.

{Since the overall SME effect is to decrease the linear polarization, a~particular measurement ($\Pi_m$) is considered compatible with a particular SME configuration if it is smaller than the SME prediction ($\Pi_m<\Pi_0$).} Therefore, the~probability of compatibility with the SME prediction $\Pi_0$ may be written as follows:
\begin{equation}
    \mathcal{P}_\mathrm{lin}=\int_0^{\Pi_0} P(\Pi|\Pi_m) dp
\end{equation}
{where $P(\Pi|\Pi_m)$ is the Rice distribution from Equation~(\ref{eqA5})}. The~total compatibility of a given dataset of both circular and linear polarimetry is obtained by multiplying all individual likelihoods together:
\begin{equation}
    \label{eqtotalcomp}
    \mathcal{P}=\prod_i \mathcal{P}_\mathrm{circ}^{(i)}\mathcal{P}_\mathrm{lin}^{(i)}
\end{equation}

\noindent where the product is taken over all measured sources. The~constraints on the $10$ independent SME parameters are derived by maximizing the likelihood with respect to them and extracting the standard errors, as~detailed in Section~\ref{section:results}.

\section{Source~Parameters}\label{section:source}
The likelihood model introduced in the previous section requires some assumption of the photon polarization state at the source, $\pmb{s}_z$. In~this section, we introduce our assumptions for the polarization angle and linear and circular polarization fractions at the source: $\psi_z$, $p_z$, and $v_z$. In~general, we aim to choose the most conservative values, i.e.,~values for which 
fewest SME configurations are ruled out or, alternatively, values that maximize the compatibility likelihood $\mathcal{P}$. {By maximizing $\mathcal{P}$, we seek the weakest possible constraints on LIV supported by the data. We emphasize that weaker constraints do not imply that a future positive detection is more likely. Instead, they merely define the region of parameter space that we can claim to have thoroughly explored with high confidence.}

\newpage

\subsection{Circular~Polarization}

For all observed astronomical sources, we assume the circular polarization at the source to be zero ($v_z=0$). This assumption is justified by both experimental and theoretical studies in 
literature~\cite{circ_zero_1,circ_zero_2,circ_zero_3,circ_zero_4} that found negligible or very small circular polarization for the majority of extragalactic sources. For~sources with small but non-negligible circular polarization ($0<|v_z|\ll p_z$), the~assumption of $v_z=0$ remains conservative, since non-zero $v_z$ results in an increase in the magnitude of observed circular polarization (Equation \ref{eq24}) and, therefore, weaker constraints on SME coefficients (larger $\mathcal{P}_\mathrm{circ}$).

\subsection{Linear~Polarization}

In our previous works~\cite{kislat17,kislat18,andy}, large values of $p_z$ were adopted as 
most conservative since $\mathcal{P}_\mathrm{lin}$ increases with $p_z$ in the limit of weak SME (small $\Phi$) and the dominance of linear polarization over circular ($|v_z|\ll p_z$). Specifically, the~most conservative assumption for the value of $p_z$ was taken as the largest possible source polarization fraction. For~example, \mbox{the authors in \cite{kislat17,kislat18}} adopted $p_z=1.0$ for all sources ignoring astrophysical considerations, while~\cite{andy} adopted $p_z=0.7$ based on the literature polarimetry of low redshift sources, resulting in somewhat stricter constraints on SME~parameters.

\begin{figure}[H]
\includegraphics[width=\columnwidth]{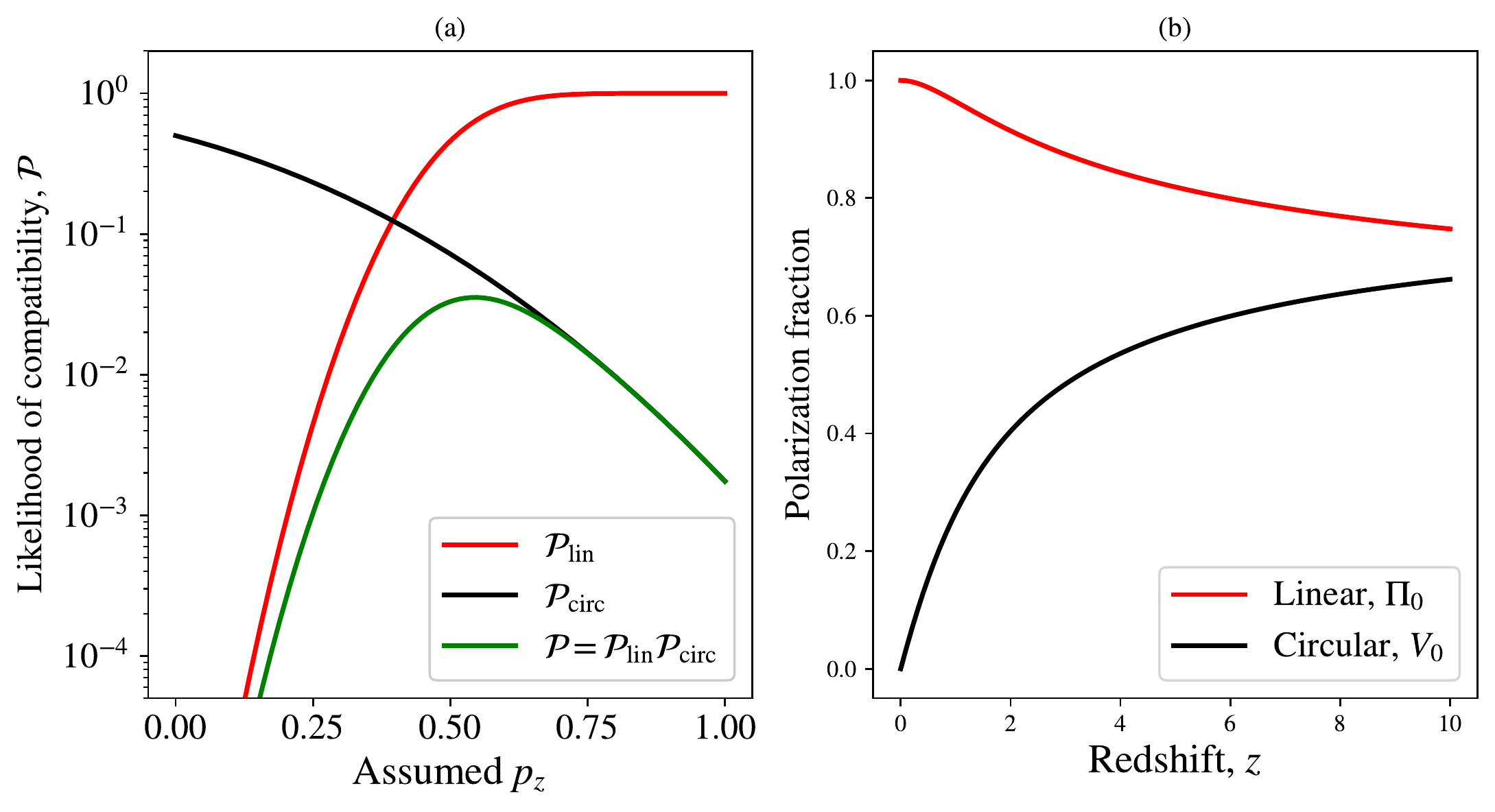}
\caption{(\textbf{a}) Plots of the linear, circular, and total likelihood of compatibility for a single test measurement as a function of the assumed linear polarization fraction at the source. The~astrophysical source is assumed to be located at the Vernal equinox ($\alpha=0$, $\delta=0$) and $z=2.0$. For~demonstration purposes, the~adopted test measurement is $\Pi_m=0.5\pm 0.1$, $V_m=0.00\pm0.01$ and $\Psi_m=0.0$. The~adopted SME configuration has all ten parameters considered in this study set to $10^{-34}$ (the order of magnitude of the upper limit derived in~\cite{andy,kislat18}). The~test measurement is assumed to have been taken through the EFOSC2~\cite{EFOSC} \textit{V} filter in zenith (see Section~\ref{section:dataset}). The linear compatibility increases while the circular compatibility decreases with $p_z$. The~most conservative assumption for the value of $p_z$ is the one maximizing the total compatibility, which, in this case, occurs around $p_z=0.55$. (\textbf{b}) The effect of SME on the linear and circular polarization fractions, shown here as $\Pi_0$ and $V_0$ as functions of the redshift of the source. The~predictions shown here were derived for a test source at the Vernal equinox ($\alpha=0$, $\delta=0$) observed through the EFOSC2~\cite{EFOSC} \textit{V} filter in zenith. The~initial polarization state of the photons was assumed to be $v_z=0$, $p_z=1$, and $\psi_z=-30^\circ$. The~adopted SME configuration is the same as in (\textbf{a}).}
\label{fig:maximization}
\end{figure}

However, previous work disregarded the fact that $\mathcal{P}_\mathrm{circ}$, in~general, decreases with $p_z$ since larger linear polarization at the source results in the faster conversion of linear to circular polarization in-flight and, hence, larger observed circular polarization $V_0$. Therefore, the~assumption of large $p_z$ may not be 
most conservative in certain instances when circular polarimetry is available. An~example of this effect is demonstrated in Figure~\ref{fig:maximization}a. 

In~the figure, the~values of $\mathcal{P}_\mathrm{lin}$, $\mathcal{P}_\mathrm{circ}$, and the total compatibility $\mathcal{P}$ are shown for a test observation and a test SME configuration as a function of the assumed source polarization $p_z$. $\mathcal{P}_\mathrm{lin}$ increases with $p_z$ monotonically with an asymptote at $\mathcal{P}_\mathrm{lin}=1$, suggesting higher values of $p_z$ as 
most conservative in the absence of other considerations. On~the other hand, $\mathcal{P}_\mathrm{circ}$ decreases with $p_z$ monotonically. Therefore, a~more conservative assumption for $p_z$ is one that maximizes the total compatibility, which, in this example, falls around $p_z=0.55$.

In this study, we determined the most conservative value of $p_z$ by numerically maximizing the total likelihood of compatibility, $\mathcal{P}_\mathrm{lin}\mathcal{P}_\mathrm{circ}$, for~each available polarization~measurement.

\subsection{Polarization~Angle}

The polarization angle at the source, $\psi_z$, may be estimated by seeking a value that reproduces the observed polarization angle, $\Psi_0=\Psi_m$, under~the most conservative assumptions for $p_z$ and $v_z$ as described above. Combining Equations~(\ref{eq22})--(\ref{eq24}) and solving for $\psi_z$:
\begin{equation}
    \psi_z=\psi_z'+\frac{\xi}{2}=\frac{1}{2} \mathrm{atan2}\left(\frac{U_m'}{\mathcal{T}[\cos(2\Phi)]},Q_m'\right)+\frac{\xi}{2}
\end{equation}

\noindent where $Q_m'=\Pi_m' \cos(2\Psi_m')$, $U_m'=\Pi_m' \sin(2\Psi_m')$, and $\Psi_m'=\Psi_m-\xi/2$. As~in~\cite{andy,kislat18}, the~uncertainty in $\Psi_m$ is not required by the~method.

\section{Sample~Dataset}\label{section:dataset}
We derive constraints on the 10 real SME parameters by maximizing the total likelihood of compatibility (Equation \ref{eqtotalcomp}) for a set of linear and circular polarimetric measurements from 
literature. The~dataset used in this study includes 21 quasars and was adopted from the compilation of archival linear polarimetry and original circular polarimetry in~\cite{circ_zero_1}. All data used in this analysis are reproduced in Table~\ref{pol-table}.
\startlandscape
\begin{specialtable}[H]
\caption{Coordinates, redshifts, and~optical linear/circular polarization measurements of the selected quasars. References (z Ref.) are provided for the redshift values. Transmission bands and references (p Ref.) are provided for the linear polarization measurements. All circular polarization measurements were taken through the Bessel \textit{V} filter and were reported in \citep{circ_zero_1}. The~detection efficiency profiles for the Bessel \textit{V} filter, Ga-As photomultiplier, and~Na-K-Cs-Sb (S20) photomultiplier are shown in Figure~\ref{fig:filters}.}
\widetable
\setlength{\tabcolsep}{5.45mm}
\begin{tabular}{cccccccccc}
\toprule
\textbf{Object Identifier} & \textbf{RAJ2000 ($\alpha$)} & \textbf{DEJ2000 ($\delta$)} & \textbf{z} & \textbf{z Ref.} & $\mathbf{\Pi_m~(\boldsymbol{\%})}$ &  $\mathbf{\Psi_m~(\boldsymbol{^\circ})}$ & \textbf{Band} & \textbf{p Ref.} & $\mathbf{V_m~(\boldsymbol{\%})}$\\
\midrule
QSO B1120+0154 & $11~23~20.73$ & $+01~37~47.5$ & $1.47$ & \citep{SDSS_dr7} & $1.95 \pm 0.27$ & $9 \pm 4$ & V & \citep{hutsemekers_1998} & $-0.02 \pm 0.05$ \\ 
QSO B1124-186 & $11~27~04.39$ & $-18~57~17.4$ & $1.05$ & \citep{VLBA} & $11.68 \pm 0.36$ & $37 \pm 1$ & V & \citep{sluse_2005} & $-0.04 \pm 0.08$ \\
QSO J1130-1449 & $11~30~07.05$ & $-14~49~27.4$ & $1.19$ & \citep{6dFGS_dr3} & $1.30 \pm 0.40$ & $23 \pm 10$ & Ga-As & \citep{impey_1990} & $-0.05 \pm 0.05$ \\
QSO B1157+014 & $11~59~44.83$ & $+01~12~07.0$ & $2.00$ & \citep{KODIAQ_dr2} & $0.76 \pm 0.18$ & $39 \pm 7$ & V & \citep{lamy_2000} & $-0.10 \pm 0.08$ \\
LBQS 1205+1436 & $12~08~25.38$ & $+14~19~21.1$ & $1.64$ & \citep{SDSS_dr7} & $0.83 \pm 0.18$ & $161 \pm 6$ & V & \citep{lamy_2000} & $-0.10 \pm 0.09$ \\
LBQS 1212+1445 & $12~14~40.27$ & $+14~28~59.3$ & $1.63$ & \citep{SDSS_dr9} & $1.45 \pm 0.30$ & $24 \pm 6$ & V & \citep{hutsemekers_1998} & $0.15 \pm 0.09$ \\
QSO B1215-002 & $12~17~58.73$ & $-00~29~46.3$ & $0.42$ & \citep{SDSS_dr9} & $23.94 \pm 0.70$ & $91 \pm 1$ & V & \citep{sluse_2005} & $-0.42 \pm 0.40$ \\
QSO B1216-010 & $12~18~34.93$ & $-01~19~54.3$ & $0.554$ & \citep{veron} & $11.20 \pm 0.17$ & $100 \pm 1$ & V & \citep{sluse_2005} & $-0.01 \pm 0.07$ \\
Ton 1530 & $12~25~27.40$ & $+22~35~13.0$ & $2.05$ & \citep{SDSS_dr14} & $0.92 \pm 0.14$ & $169 \pm 4$ & V & \citep{sluse_2005} & $0.01 \pm 0.10$ \\
QSO J1246-2547 & $12~46~46.80$ & $-25~47~49.3$ & $0.63$ & \citep{VLBA} & $8.40 \pm 0.20$ & $110 \pm 1$ & Ga-As & \citep{impey_1990} & $-0.23 \pm 0.20$ \\
QSO B1246-0542 & $12~49~13.86$ & $-05~59~19.1$ & $2.23$ & \citep{KODIAQ_dr2} & $1.96 \pm 0.18$ & $149 \pm 3$ & Ga-As & \citep{schmidt_1999} & $0.01 \pm 0.03$ \\
QSO B1254+0443 & $12~56~59.92$ & $+04~27~34.4$ & $1.02$ & \citep{SDSS_dr9} & $1.22 \pm 0.15$ & $165 \pm 3$ & Ga-As & \citep{berriman_1990} & $-0.02 \pm 0.04$ \\
QSO B1256-229 & $12~59~08.46$ & $-23~10~38.7$ & $0.481$ & \citep{ESO_VLT} & $22.32 \pm 0.15$ & $157 \pm 1$ & V & \citep{sluse_2005} & $0.18 \pm 0.04$ \\
QSO J1311-0552 & $13~11~36.56$ & $-05~52~38.6$ & $2.19$ & \citep{2MASS} & $0.78 \pm 0.28$ & $179 \pm 11$ & V & \citep{hutsemekers_1998} & $-0.08 \pm 0.06$ \\
LBQS 1331-0108 & $13~34~28.06$ & $-01~23~49.0$ & $1.78$ & \citep{SDSS_dr9} & $1.88 \pm 0.31$ & $29 \pm 5$ & V & \citep{hutsemekers_1998} & $-0.04 \pm 0.06$ \\
{[VV96]} J134204.4-181801 & $13~42~04.41$ & $-18~18~02.6$ & $2.21$ & \citep{2MASS} & $0.83 \pm 0.15$ & $20 \pm 5$ & V & \citep{sluse_2005} & $-0.01 \pm 0.07$ \\
2E 3238 & $14~19~03.82$ & $-13~10~44.8$ & $0.13$ & \citep{INTEGRAL} & $1.63 \pm 0.15$ & $44 \pm 3$ & Ga-As & \citep{berriman_1990} & $0.05 \pm 0.06$ \\
LBQS 1429-0053 & $14~32~29.25$ & $-01~06~16.1$ & $2.08$ & \citep{SDSS_dr9} & $1.00 \pm 0.29$ & $9 \pm 9$ & V & \citep{hutsemekers_1998} & $0.02 \pm 0.08$ \\
QSO J2123+0535 & $21~23~44.52$ & $+05~35~22.1$ & $1.88$ & \citep{JVAS} & $10.70 \pm 2.90$ & $68 \pm 6$ & Ga-As & \citep{impey_1990} & $0.02 \pm 0.15$ \\
QSO B2128-123 & $21~31~35.26$ & $-12~07~04.8$ & $0.50$ & \citep{KODIAQ_dr2} & $1.90 \pm 0.40$ & $64 \pm 6$ & S20 & \citep{visvanathan_1998} & $-0.04 \pm 0.03$ \\
QSO B2155-152 & $21~58~06.28$ & $-15~01~09.3$ & $0.67$ & \citep{VLBA} & $22.60 \pm 1.10$ & $7 \pm 2$ & Ga-As & \citep{impey_1990} & $-0.35 \pm 0.10$ \\
\bottomrule
\end{tabular}
\label{pol-table}

\end{specialtable}
\finishlandscape

Circular polarization measurements in~\cite{circ_zero_1} were obtained through the Bessel \textit{V} filter (ESO \#641) using the EFOSC2~\cite{EFOSC} instrument mounted on the ESO 3.6 m telescope at La Silla~Observatory.

The linear polarization measurements for the same sources were compiled by~\cite{circ_zero_1} from seven references that we refer to as Impey+1990~\cite{impey_1990}, Berriman+1990~\cite{berriman_1990}, Hutsem{\'e}kers+1998~\cite{hutsemekers_1998}, Visvanathan+1998~\cite{visvanathan_1998}, Schmidt+1999~\cite{schmidt_1999}, Lamy+2000~\cite{lamy_2000}, and Sluse+2005~\cite{sluse_2005}.

The measurements from Impey+1990~\cite{impey_1990} were obtained with an unfiltered Ga-As photomultiplier tube in the MINIPOL polarimeter mounted on the Irénée du Pont 100 inch telescope at the Las Campanas Observatory. The~measurements from Berriman+1990~\cite{berriman_1990} were obtained with an unfiltered Ga-As photomultiplier in the Two-Holer polarimeter/photometer mounted on the Bok 2.3 m telescope at the Steward Observatory and the 1.5 m telescope at the Mount Lemmon Observing Facility. The~measurements from Schmidt+1999~\cite{schmidt_1999} were obtained unfiltered using polarimeters mounted on the \mbox{Bok 2.3 m} telescope at the Steward Observatory. The~publication does not specify the detector; however, it is mentioned that the waveband and sensitivity were very similar to the Breger polarimeter~\cite{breger_1977} at the McDonald Observatory. 

Since~\cite{breger_1977} employ the response curve of a Ga-As photomultiplier tube in their analysis, we assume that the Steward Observatory polarimeters used some type of Ga-As tubes as well. The~measurements from Visvanathan+1998~\cite{visvanathan_1998} were obtained with an unfiltered Na-K-Cs-Sb (S20) photomultiplier in an automated polarimeter~\cite{visvanathan_1972} mounted on the Anglo-Australian Telescope at the Siding Spring Observatory. The~measurements from Hutsem{\'e}kers+1998~\cite{hutsemekers_1998}, Lamy+2000~\cite{lamy_2000}, and Sluse+2005~\cite{sluse_2005} were obtained through the Bessel \textit{V} filter using the EFOSC1/EFOSC2~\cite{EFOSC} instrument at the La Silla Observatory, which we assume to have a sufficiently similar transmission profile to the Bessel V filter (ESO \#641) used by~\cite{circ_zero_1} for circular~polarimetry.

\subsection*{Detection Efficiency~Profiles}

Our method requires efficiency profiles, $\tau(\omega_0)$, for~each measurement in the dataset. For~filtered observations, the~transmission profile of the filter is typically self-sufficient with the atmospheric transmission, detector response, mirror reflectivity, etc. only contributing minor higher order corrections. For~unfiltered observations, the~response curve of the detector becomes the most determining factor, with the atmospheric transmission making a significant contribution at the blue end of the visible spectrum where the ozone layer predominantly sets the short wavelength cut-off of the~instrument.

For all filtered observations through the Bessel \textit{V} band, we adopted the ESO \#641 transmission profile available online 
{(\url{https://www.eso.org/sci/facilities/lasilla/instruments/efosc/inst/Efosc2Filters.html}, accessed ``13 May 2021'')}. For~unfiltered observations, we took the generic response curves of the corresponding photomultiplier tubes (Ga-As and S20) from (\cite{ebdon_1998}, p.101). All profiles were multiplied by the atmospheric transmission, based on the atmospheric radiation model in~\cite{model_atmosphere} calculated for the geographic altitude of $2400\ \mathrm{m}$ and weather conditions of Cerro Paranal. The~model employs the radiative transfer routine in~\cite{rad_trans} and HITRAN opacity database~\cite{HITRAN}. The~altitude and conditions adopted by the model are generally similar to those at the majority of aforementioned facilities contributing data to this~study.

\begin{figure}[H]
\includegraphics[width=\columnwidth]{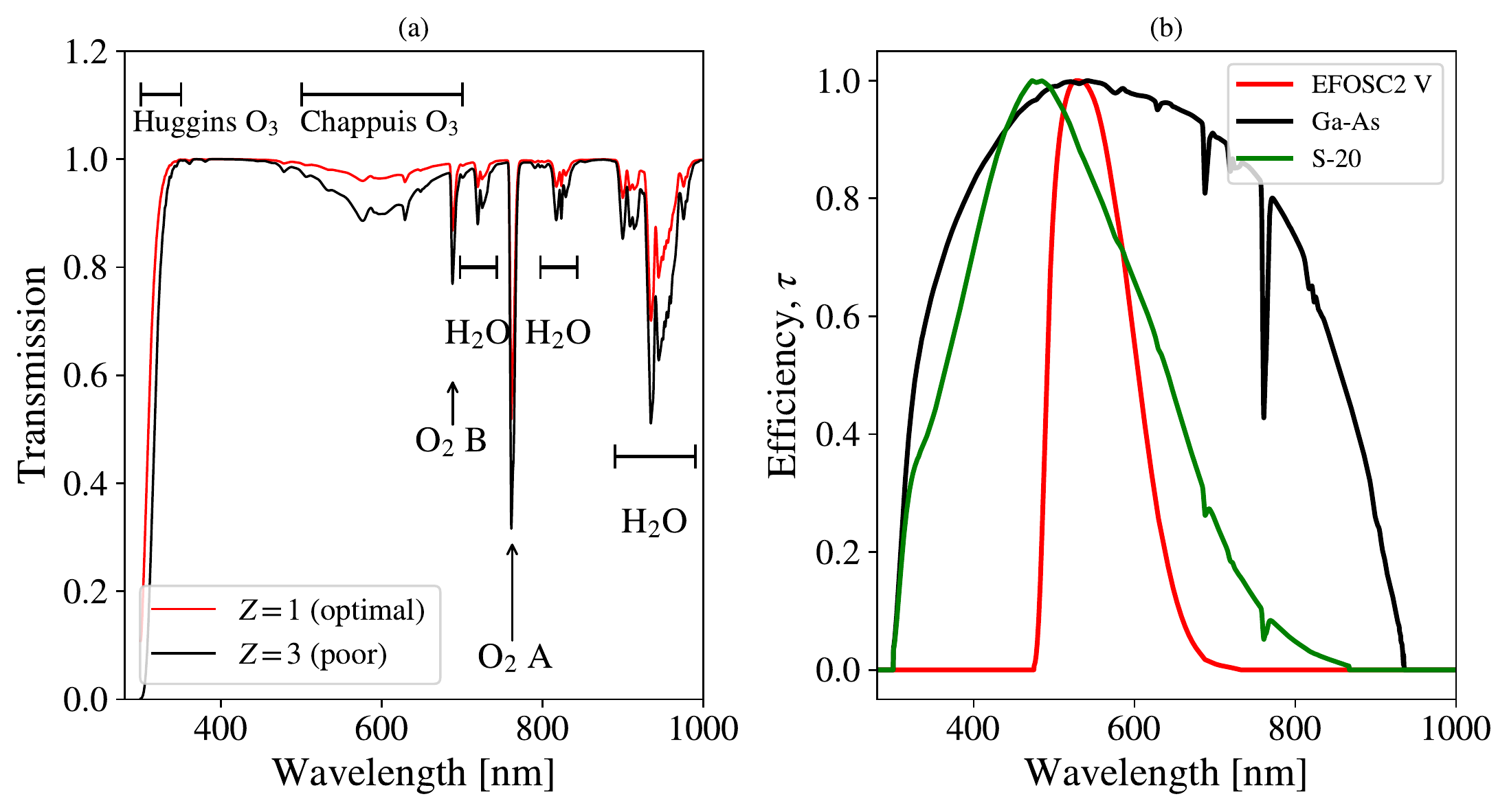}
\caption{(\textbf{a}) The atmospheric transmission model employed in this study as a function of wavelength for two airmasses: $Z=1,3$. The most prominent absorption features due to ozone ($\mathrm{O}_3$), oxygen ($\mathrm{O}_2$), and water vapor ($\mathrm{H}_2\mathrm{O}$) are labeled. For~clarity, the~plots exclude Rayleigh and Mie scattering effects (however, they are included in the analysis). (\textbf{b}) Detection efficiency profiles of the three instruments relevant to this study. For~the \textit{V} band, the~transmission of the filter is shown. For~both Ga-As and S-20 photomultipliers, the~response curves of the detector are used instead. All curves are corrected for the atmospheric transmission. In~the figure, the~$Z=1$ atmosphere was~applied.}
\label{fig:filters}
\end{figure}

Since the atmospheric transmission strongly depends on the target elevation above the horizon, we performed all calculations for the cases of both optimal (airmass $Z=1.0$) and poor (airmass $Z=3.0$) conditions. 
{In this analysis, we define the airmass, $Z$, as~the integrated atmospheric density along the line of sight to the source normalized to $Z=1$ in zenith.} 
The~final efficiency profiles used in both cases are provided in Figure~\ref{fig:filters}. All profiles are normalized to $\tau=1$ at the peak efficiency for convenience, since our method is not sensitive to multiplicative pre-factors, as~evident from Equation~(\ref{eq20}).

\section{SME~Constraints}\label{section:results}
Following~\cite{andy,kislat18}, we explored the ten-dimensional likelihood distribution given by Equation~(\ref{eqtotalcomp}) using the Metropolis--Hastings Markov-Chain Monte Carlo (MCMC) approach {implemented in \texttt{Python} package \texttt{emcee} \cite{emcee}}. Exploration was carried out by so-called \textit{walkers} that spawn at some location in the likelihood space (i.e., some combination of SME parameters) and proceed in random Monte Carlo steps. In~each step, offsets for the positions of the walkers were drawn from a suitable proposal distribution in all ten dimensions, and the likelihood of compatibility at the new location was estimated. 

The~ratio of the new likelihood to the old was then compared to a uniform random number between $0$ and $1$, and the step was accepted if the former exceeded the latter. The~non-zero probability of stepping into a lower likelihood was implemented to prevent walkers from becoming ``stuck'' in local minima if any happened to be present in the likelihood space. The~probability distribution for each of the SME parameters was then extracted by assuming the probability of each value along a given dimension to be proportional to the number of steps that the walkers spent in its immediate~proximity.

In this study, we followed~\cite{andy,kislat18} and used a Gaussian proposal distribution with the same standard deviation in all dimensions. The~standard deviation was chosen by first assuming all SME parameters to be equal and searching a value where the likelihood was evaluated to the midpoint between the extreme cases of infinitely strong ($k_{(E,B)l,m}\rightarrow \infty$) and infinitely weak ($k_{(E,B)l,m}\rightarrow 0$) SME. The~result, $\sim 2\times10^{-36}$, was expected to be of comparable magnitude to the average width of the likelihood distribution across all dimensions. The~initial positions for $20$ MCMC walkers were drawn from the proposal distribution as well. 

\begin{figure}[H]
\includegraphics[width=\columnwidth]{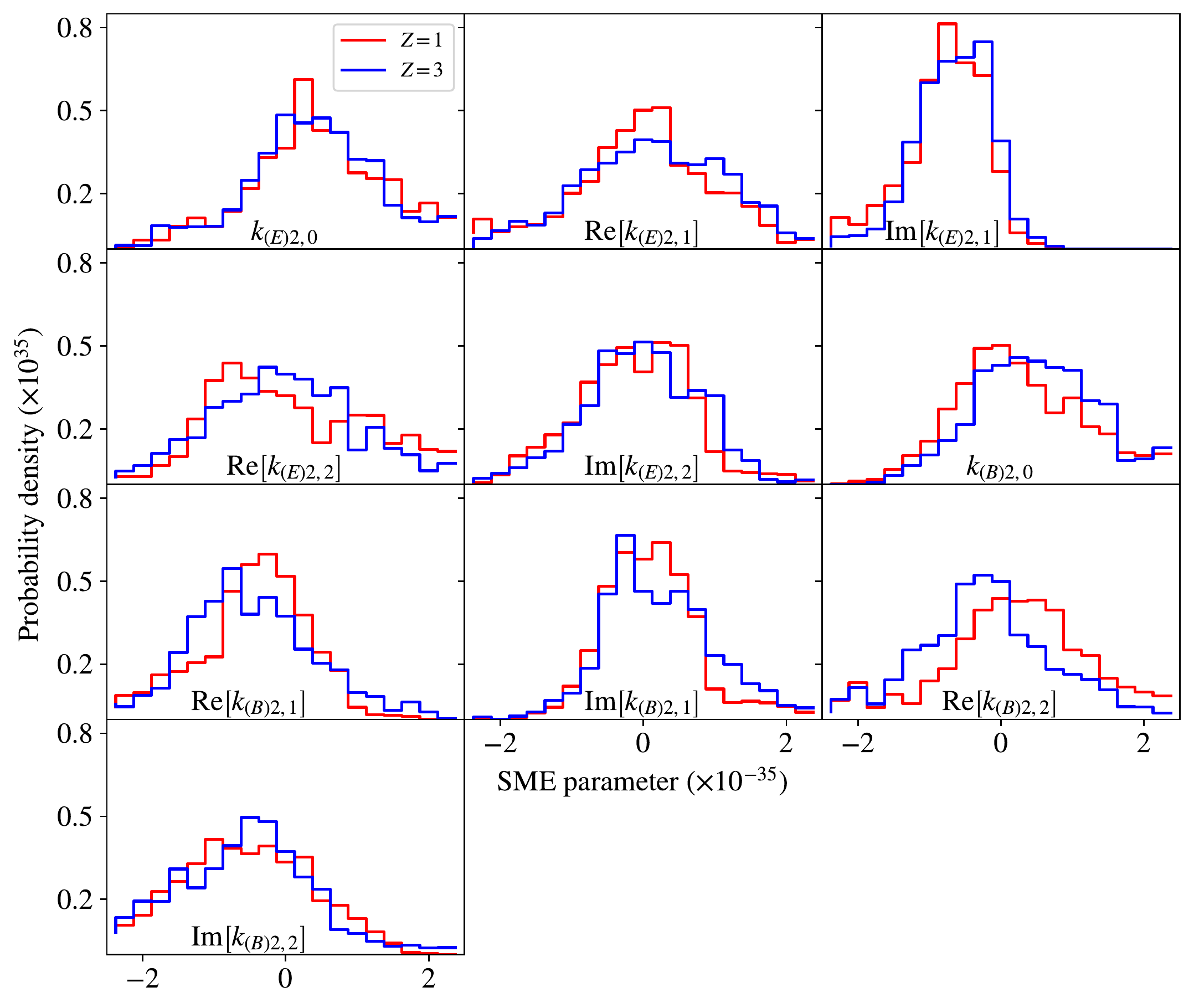}
\caption{Probability distributions for the values of the 10 real $d=4$ SME parameters extracted from MCMC chains for the dataset of optical circular and linear polarimetry considered in this study. All parameters are dimensionless. Color-coded are the cases of $Z=1$ (red) and $Z=3$ (blue) atmospheres (Section \ref{section:dataset}). The~distributions are uniformly binned with the bin width of $2.5\times10^{-36}$ for the total of $20$ bins.}
\label{fig:mcmc}
\end{figure}

The~run was terminated when each walker completed $500$ steps (accepted or rejected) for a total of $10^4$ steps across all walkers. At~the end of the chain, the~step acceptance ratios were found to be $0.21$ and $0.23$ for the $Z=1$ and $Z=3$ cases. The~extracted probability distributions for all $10$ SME parameters are given in \mbox{Figure~\ref{fig:mcmc}}. The~constraints on the individual parameters (upper and lower bounds) were taken as the 5th and 95th percentiles of the corresponding distributions. The~numeric values are summarized in Table~\ref{table:mcmc}. In~the case of differing results for the $Z=1$ and $Z=3$ cases, the~most conservative case was taken (i.e., the minimum value for the lower bound and maximum value for the upper bound).

\begin{specialtable}[H] 
\caption{The derived constraints on the values of the 10 real $d=4$ SME parameters. The~upper and lower bounds were taken as the 5th and 95th percentiles of the distributions shown in Figure~\ref{fig:mcmc}. For~each parameter, the~most conservative of the $Z=1$ and $Z=3$ cases is~shown.}
\setlength{\tabcolsep}{6.7mm}
\begin{tabular}{ccc}
\toprule
\textbf{SME Parameter} & \textbf{Upper Bound ($\times 10^{-35}$)} & \textbf{Lower Bound ($\times 10^{-35}$)}\\
\midrule
$k_{{(E)2,0}}$ &$\leq 2.9 $&$\geq -1.2 $\\
$\mathrm{Re}\left[k_{{(E)2,1}}\right]$ &$\leq 1.8 $&$\geq -1.5 $\\
$\mathrm{Im}\left[k_{{(E)2,1}}\right]$ &$\leq 0.2 $&$\geq -1.4 $\\
$\mathrm{Re}\left[k_{{(E)2,2}}\right]$ &$\leq 3.0 $&$\geq -1.7 $\\
$\mathrm{Im}\left[k_{{(E)2,2}}\right]$ &$\leq 1.4 $&$\geq -1.4 $\\
$k_{{(B)2,0}}$ &$\leq 3.2 $&$\geq -0.7 $\\
$\mathrm{Re}\left[k_{{(B)2,1}}\right]$ &$\leq 1.3 $&$\geq -1.8 $\\
$\mathrm{Im}\left[k_{{(B)2,1}}\right]$ &$\leq 1.9 $&$\geq -0.8 $\\
$\mathrm{Re}\left[k_{{(B)2,2}}\right]$ &$\leq 2.1 $&$\geq -2.1 $\\
$\mathrm{Im}\left[k_{{(B)2,2}}\right]$ &$\leq 1.2 $&$\geq -2.3 $\\
\bottomrule
\end{tabular}
\label{table:mcmc}
\end{specialtable}
\unskip

\section{Conclusions}
\label{section:conclusion}
In this work, we used optical circular and linear polarimetry of high redshift extragalactic sources to derive new constraints on the Lorentz invariance violation in the framework of the Standard-Model Extension at mass dimension $d=4$. The~numerical values of the constraints are expressed as upper and lower bounds on the 10 SME parameters summarized in Table~\ref{table:mcmc}. The~detailed probability distributions for each parameter were also calculated and are presented in Figure~\ref{fig:mcmc}. Since the exact conditions of the literature observations employed in this study are unknown, all calculations were performed for two different airmasses ($Z=1$ and $Z=3$) with no significant difference in the~results.

{Our results assume that SME effects at mass dimension $d=4$ dominate. The~focus on this particular mass dimension is justified by the fact that it represents the leading-order terms that can be tested with our method. Furthermore, the~reliance of our method on helicity coupling, which only occurs at even $d$, and~the large number of independent SME parameters at $d\geq6$ exceeding the number of lines of sight available in our dataset ($21$) prevented application of our dataset to higher mass dimensions. 
	
	For~example, $d=6$ analysis requires at least $42$ lines of sight, $d=8$ requires $90$ lines of sight, etc. Nonetheless, the~method proposed in this work can be extended to those mass dimensions given a larger number of measurements. A~follow-up study of higher mass dimensions would ideally also use higher-energy data due to the stronger suppression of these terms. However, circular polarization measurements of astrophysical X-rays are currently unavailable.}

{In our analysis, we assumed the dominance of linear polarization over circular at all sources as well as sufficiently weak LIV such that said dominance was maintained throughout the line of sight. Both assumptions are well justified by theoretical and experimental astrophysical considerations~\cite{circ_zero_1,circ_zero_2,circ_zero_3,circ_zero_4} as well as existing SME constraints~\cite{andy,kislat18,datatables}. Due to the difficulty of determining the initial polarization state of photons at the source, we searched the space of all possible initial polarization states numerically and assumed the one that resulted in the most conservative SME constraints for any given source. Finally, we assumed a weak wavelength dependence of polarization at the source across the optical range justified by the nearly flat polarization spectra of nearby active galactic nuclei as reported in our previous work~\cite{kislat17}.}

Similarly to~\cite{andy,kislat18,datatables}, we found the probability distributions for individual SME parameters (Figure \ref{fig:mcmc}) were consistent with zero within two sigma. Unlike~\cite{andy,kislat18,datatables}, our results display more asymmetry around the origin, with the most notable example being $\mathrm{Im}\left[k_{(E)2,1} \right]$, which appears heavily skewed toward negative values. We note that circular polarimetry is generally expected to be more sensitive to the sign of coefficients than linear polarimetry due to the fact that both clockwise and anticlockwise circular polarization may be induced by the SME. The~observed asymmetry in the distribution is likely caused by the particular combination of clockwise/anticlockwise circular polarization measurements along the lines of sight considered in the sample~dataset.

The discussion of systematic errors in~\cite{andy} is mostly applicable to this study as well. In~our method, it is assumed that any drift in the photon polarization state between the source and the telescope is induced by the LIV and not by astrophysical processes. We first note that any process that reduces linear or enhances circular polarization would weaken our constraints due to the conservative likelihood model used. 

While modelling such effects may improve the derived constraints further, we are at no risk of falsely ruling out viable SME configurations. If~unaccounted, said astrophysical processes may eventually establish a lower bound on the parameter constraints derivable using our method. Given the considerable improvement of the constraints derived in this study compared to its predecessors (see below), it is reasonable to assume that such a lower bound has not yet been~reached.

On the other hand, astrophysical processes that enhance linear or reduce circular polarization are of much greater concern as they would falsely tighten our constraints on SME effects. Fortunately, very few such processes are known to occur in the optical regime, and most are expected to average out over multiple lines of sight. A~prominent example of such an effect includes the polarization and de-polarization of radiation by the interstellar medium~\cite{dustpol1,dustpol2}.

Our constraints are directly comparable to those derived in~\cite{andy,kislat18} {(also listed in~\cite{datatables})} and are tighter than both by approximately an order of magnitude due to our use of circular polarimetry, which has not been previously applied in similar studies. We note further that this result was achieved with only $21$ unique lines of sight, which is less than the dataset in~\cite{kislat18} containing $27$ lines of sight and much less than the dataset in~\cite{andy} containing \mbox{$1278$ lines} of sight. Our analysis demonstrates that deriving constraints from circular and linear polarimetry simultaneously is a more efficient method than the methods previously employed in literature as not only does it provide better constraints for fewer sources but it~is also free of the somewhat arbitrary assumption of an excessively high initial linear polarization fraction. 

We expect that a larger sample size as well as higher quality circular polarimetry may significantly improve the constraints derived in this work. Unfortunately, the~scarcity of circular measurements for extragalactic sources, in~part due to their characteristically weak signal, imposes strict limits on the maximum improvement one may expect from archival data and calls for new optical polarization surveys. Additionally, the~process of estimating the initial linear polarization at the source, $p_z$, through numerical optimization is far more computationally demanding than the methods in~\cite{andy,kislat18}, limiting the number of MCMC steps used in this study ($10^4$ compared to $0.5\times 10^6$ in~\cite{andy} and $10^7$ in~\cite{kislat18}) and potentially requiring high performance computing for the adequate processing of a larger dataset in the~future.

\authorcontributions{R.G. led the analysis of linear and circular polarization data. P.B. collected the dataset of linear and circular polarization data, filter transmission curves, and PMT efficiencies. F.K. co-developed the analysis~method. All authors have read and agreed to the published version of the manuscript.}

\funding{FK acknowledges NASA support under grant~80NSSC18K0264.}

\acknowledgments{The authors would like to thank Alan Kosteleck\'y, Matthew Mewes, David Mattingly, Henric Krawczynski, Jim Buckley, Floyd Stecker, and~Brian Keating for fruitful discussions. We are grateful to the late Andy Friedman without whom this collaboration would never have happened. This work made use of the SIMBAD database, operated at CDS, Strasbourg, France.}

\conflictsofinterest{The authors declare no conflict of interest.
} 

\appendixtitles{yes} 
\appendixstart
\appendix
\section{Polarization~Statistics}\label{section:appendix}
In this appendix, we present our treatment of the linear polarization degree distribution and the linear polarization bias. {Our argument is an adaptation of similar treatments in the context of X-Ray polarimetry~\cite{pol_stats_xray} and radio interferometry~\cite{pol_stats_radio} to the optical regime.} Assuming no variability in the source, there exists some ``true'' linear polarization degree, $\hat{p}$, which is related to the ``true'' intensity-normalized Stokes $\hat{q}$ and $\hat{u}$ parameters:
\begin{equation}
    \hat{p}=\sqrt{\hat{q}^2+\hat{u}^2}.
\end{equation}

We assume the measured Stokes parameters, $q$ and $u$, to~be normally distributed around their ``true'' counterparts, $\hat{q}$ and $\hat{u}$, with~identical standard deviations of $\sigma$. Furthermore, we assume no correlation between $q$ and $u$. The~probability density of observing a particular pair of $q$ and $u$ is then a product of their respective Gaussian distributions:
\begin{equation}
    P(q,u|\hat{q},\hat{u})=\frac{1}{2\pi\sigma^2} \exp\left( -\frac{(q-\hat{q})^2}{2\sigma^2} \right) \exp\left( -\frac{(u-\hat{u})^2}{2\sigma^2} \right).
\end{equation}

Equivalently, in~polar coordinates:
\begin{equation}
    P(p,\psi|\hat{p},\hat{\psi})=\frac{p}{\pi\sigma^2} \exp\left(-\frac{1}{2\sigma^2}\left[p^2+\hat{p}^2-2p\hat{p}\cos(2\psi-2\hat{\psi})\right]\right)
    \label{eqA3}
\end{equation}

\noindent where we substituted $q=p \cos(2\psi)$, $u=p \sin(2\psi)$ and their ``true'' counterparts. Note the added pre-factor of $2p$ due to the coordinate transformation, $dqdu=2pdpd\psi$. Now integrate Equation~(\ref{eqA3}) over all $\psi$ to obtain the distribution of $p$ independently of $\psi$:
\begin{equation}
    P(p|\hat{p})=\int_0^\pi P(p,\psi|\hat{p},\hat{\psi}) dp d\psi=\frac{p}{\sigma^2} \exp\left( -\frac{p^2+\hat{p}^2}{2\sigma^2} \right) I_0\left( \frac{p \hat{p}}{\sigma^2} \right)
    \label{eqA4}
\end{equation}

\noindent where $I_n(x)$ is the modified Bessel function of the first kind. 
Rewriting Equation~\mbox{(\ref{eqA4})} in terms of the exponentially scaled Bessel function, $I_0(x)=\exp(|x|)i_0(x)$, alleviates numerical issues at small $\sigma$ due to diverging $I_0(x)$:
\begin{equation}
    P(p|\hat{p})=\frac{p}{\sigma^2} \exp\left( -\frac{(p-\hat{p})^2}{2\sigma^2} \right) i_0\left( \frac{p \hat{p}}{\sigma^2} \right).
    \label{eqA5}
\end{equation}

\begin{figure}[H]
\includegraphics[width=0.9\columnwidth]{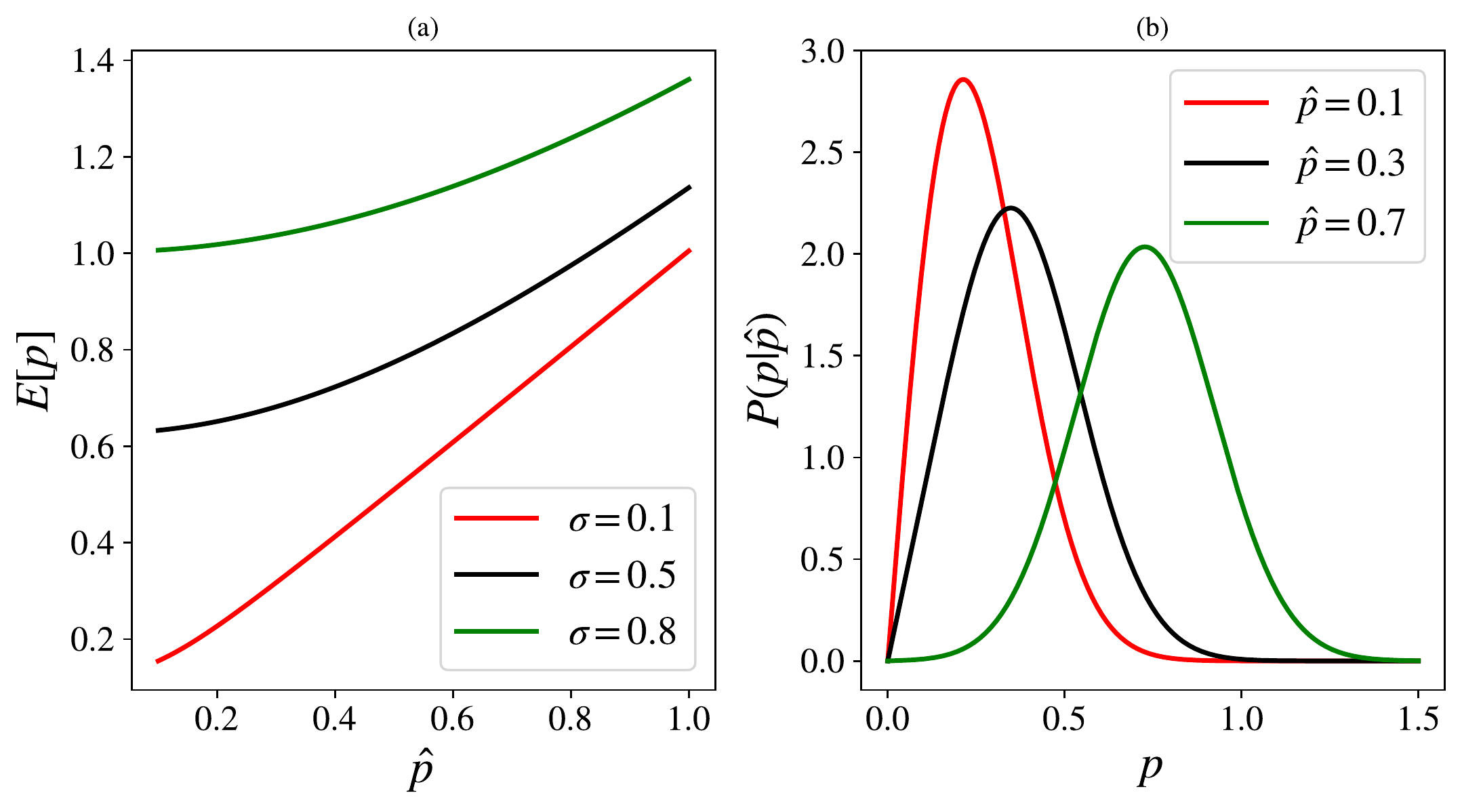}
\caption{(\textbf{a}) Plots of the expected linear polarization fraction measurement, $E[p]$ as a function of the ``true'' polarization fraction $\hat{p}$, where the expected value is taken as the mean of the corresponding statistical distribution (Equation (\ref{eqA6})). The~plots illustrate the linear polarization bias (i.e., the mismatch between expected and true values) due to $p$ being an intrinsically positive quantity. The~bias is exacerbated at larger experimental errors $\sigma$. (\textbf{b}) The probability distribution of linear polarization fraction measurements for a variety of ``true'' values $\hat{p}$ and $\sigma=0.2$, given in Equation~(\ref{eqA5}).}
\label{fig:p_bias}
\end{figure}

The distribution is plotted in Figure~\ref{fig:p_bias}. The~value of $\sigma$ in Equations~(\ref{eqA4}) and (\ref{eqA5}) may not be known {a priori}. To~address this issue, we calculate the expected value, $E[p]$, and~the variance, $\mathrm{Var}[p]$, of~$p$ using the derived distribution:
\begin{equation}
    E[p]=\int_0^\infty p P(p|\hat{p})dp=\frac{\sqrt{\pi}}{2\sqrt{2}\sigma} \exp\left( -\frac{\hat{p}^2}{4\sigma^2}\right)\left( (\hat{p}^2+2\sigma^2) I_0\left( \frac{\hat{p}^2}{4\sigma^2} \right) +\hat{p}^2 I_1\left( \frac{\hat{p}^2}{4\sigma^2} \right) \right)
    \label{eqA6}
\end{equation}
\begin{equation}
    \mathrm{Var}[p]=E[p^2]-(E[p])^2=\int_0^\infty p^2 P(p|\hat{p})dp-(E[p])^2=\hat{p}^2+2\sigma^2-(E[p])^2.
    \label{eqA7}
\end{equation}

Note that, in~general, $E[p]\neq \hat{p}$ due to the polarization degree bias. This effect is demonstrated in Figure~\ref{fig:p_bias}. For~a given polarization measurement ($E[p]$ or $\hat{p}$ depending on whether the value was debiased or not) and its uncertainty ($\sqrt{\mathrm{Var}[p]}$), one may solve Equations~(\ref{eqA6}) and (\ref{eqA7}) simultaneously for~$\sigma$.

\end{paracol}

\reftitle{References}

\end{document}